\documentclass[12pt]{iopart}

\usepackage[final]{graphicx}
\usepackage{epsf,rotating,wrapfig}

\begin{document}

\title[TOF-B$\rho$ Mass Measurements at the NSCL]{TOF-B$\rho$ Mass Measurements of Very Exotic Nuclides
for~Astrophysical~Calculations  at~the~NSCL}

\author{M~Mato\v{s}$^{1,2}$, A~Estrade$^{1,2,3}$,
M~Amthor$^{1,2,3}$, A Aprahamian$^{2,4}$, D~Bazin$^{1}$,
A~Becerril$^{1,2,3}$, T~Elliot$^{1,2,3}$, D~Galaviz$^{1,2}$,
A~Gade$^{1,3}$, S~Gupta$^{7}$, G~Lorusso$^{1,2,3}$,
F~Montes$^{1,2}$, J~Pereira$^{1,2}$, M~Portillo$^{1}$,
A~M~Rogers$^{1,2,3}$, H~Schatz$^{1,2,3}$, D~Shapira$^{5}$,
E~Smith$^{2,6}$, A~Stolz$^{1}$ and M~Wallace$^{7}$}

\address{$^{1}$ National Superconducting Cyclotron Laboratory, Michigan State University, East Lansing, MI, USA   \\
$^{2}$ Joint Institute of Nuclear Astrophysics, Michigan State University, East Lansing, MI, USA   \\
 $^{3}$ Department of Physics and Astronomy, Michigan State University, East Lansing, MI, USA   \\
 $^{4}$ Institute of Structure and Nuclear Astrophysics, Department of Physics, University of Notre Dame, South Bend, IN, USA \\
 $^{5}$ Oak Ridge National Laboratory, Oak Ridge, TN, USA   \\
 $^{6}$ Department of Physics, The Ohio State University,  Columbus, OH, USA   \\
$^{7}$ Los Alamos National Laboratory, Los Alamos, NM, USA }
\ead{matos@nscl.msu.edu}

\begin{abstract}
Atomic masses play a crucial role in many nuclear astrophysics calculations.
The lack of experimental values for relevant  exotic nuclides
 triggered a rapid development of new mass measurement devices around the world.
The Time-of-Flight (TOF) mass measurements offer a~complementary technique to the
most precise one, Penning trap measurements~\cite{blaum06}, the latter being limited by
the rate and half-lives of the~ions of interest. The
NSCL facility provides a well-suited infrastructure for TOF mass
measurements of very exotic nuclei. At this facility, we have recently implemented 
a~TOF-B$\rho$ technique and performed mass measurements of
neutron-rich nuclides in the Fe region, important for r-process
calculations and for calculations of processes occurring in the
crust of accreting neutron stars.
\end{abstract}

%Uncomment for PACS numbers title message
\pacs{21.10.Dr, 26.50.+x, 26.60.+c, 29.27.Eg}
% Keywords required only for MST, PB, PMB, PM, JOA, JOB?
%\vspace{2pc}
%\noindent{\it Keywords}: Article preparation, IOP journals
% Uncomment for Submitted to journal title message
%\submitto{\JPA}
% Comment out if separate title page not required
\maketitle

\section{Introduction}

%The TOF mass measurements offer a~complementary technique to the
%most precise Penning trap measurements, the latter being limited by
%the rate, half-lives and the chemistry of~the~ions of interest.
With~a~minimum rate requirement of the order of 0.01 particles per second
and a measurement time shorter than 1~$\mu$s, the TOF mass
measurements offer a~complementary technique to the
most precise Penning trap measurements.
%access the most exotic nuclei. 
An~entire region of the
chart of nuclides 
%that could not be reached by other techniques 
can
be covered in just one experimental run, producing important data
related to astrophysical calculations, tests of mass models, and other nuclear
physics applications.

\section{Principle of the TOF Mass Measurement Technique}
\label{sec:tofmethod}
Time-of-Flight mass measurements are based on a precise measurement
of the time it takes a charged particle to travel between two points within
a magnetic optics system with a magnetic
rigidity $B\rho$. The mass of the ion is determined from the
equation
\begin{equation}
\frac{m}{q} =  \frac{B \rho}{v} = B\rho \frac{t}{L},
\end{equation}
where $t$ is the time-of-flight of a particle with mass-to-charge ratio $m/q$ and
velocity $v$ within the path length $L$. In the relativistic case the
expression can be written as
\begin{equation}
\frac{m_0}{q} = B \rho \left( \frac{1}{v^2} - \frac{1}{c^2} \right)^{1/2} = B
\rho \left( \frac{t^2}{L^2} - \frac{1}{c^2}  \right)^{1/2},
\end{equation}
where $c$ is the speed of light.

In reality, precise measurements of the path length and absolute
measurements of magnetic rigidity are very difficult. Nevertheless, as the path
length and the magnetic rigidity can be considered to be constant
for a fixed configuration of the system, 
the direct relation between the time-of-flight and the mass-to-charge
ratio can be found using the nuclides with well known mass values.

%and nuclides with well known mass values are used
%to find the relation between the time-of-flight and the mass-to-charge
%ratio.

A~typical beam line with a momentum acceptance of several percent
does not fulfil the condition of constant B$\rho$ required for a high precision mass
measurement. To overcome this problem, two solutions are adapted at
existing TOF mass measurement facilities. A measurement of
relative magnetic rigidity by a position measurement at a dispersive
plane of the magnetic optics system is used at the SPEG spectrometer at GANIL \cite{ganil} and
at the NSCL TOF-B$\rho$ facility (discussed in this paper). These measurements have
to address the Z dependence due to energy losses in a position
sensitive detector. The other possibility, applied by TOFI at~LANL~\cite{tofi} 
and IMS at GSI~\cite{hausmann}, is the use of isochronous
mode in magnetic optics. In this case, the ions with higher velocities
travel on the longer trajectories in such a way that ions 
with the same mass-to-charge ratios
%of the same species 
have the same time-of-flight values independently from their
energies. This can be achieved for a non-relativistic case if
\begin{equation}
\frac{L}{B\rho}= \mbox{constant}
\end{equation}
for the full acceptance of the beam line. For
relativistic cases a term dependent on a mass-to-charge ratio starts
to play an important role;
\begin{equation}
L\left[\left(\frac{1}{B \rho}\right)^2 + \left(\frac{1}{c
m_0/q}\right)^2 \right]^{1/2} = \mbox{constant}
\end{equation}
and the isochronicity is limited to nuclides within a small range
of mass-to-charge ratios. 
This problem can be alleviated by an improved time-of-flight resolution
 for non-isochronous particles
with decreased magnetic rigidity acceptance~\cite{matosthesis}.
In this case, the statistical uncertainty does not have to suffer from the reduced rate 
as it is compensated by a better resolution, except
%This is not fulfilled 
for the region of the best isochronicity, where
the resolution is not improved with smaller acceptance.
%As this is usually the case for exotic nuclei of interest with low
%rates, this method can only be applied in certain cases.
%
%(usually an exotic one with a low rate) 
%the resolution is not improved with a smaller acceptance
% and it suffers from the rate reduction.
%
%By~lowering the acceptance the resolution is not improved 
%in the area of isochronicity that is
%usually also the area of interest with a low rate.
%

Measurements of
the magnetic rigidity were recently reported at GSI for the IMS technique~\cite{chen}. At the
future GSI facility a pair of time-of-flight detectors allowing a
velocity measurement has been proposed \cite{matoscgs}.

All the mentioned techniques  are one-turn
time-of-flight facilities except for the IMS, which  uses a storage ring for
multiple TOF measurements. 
This leads to a significantly better resolution and is important for resolving
isomeric contaminants in the TOF spectra.

\section{TOF-B$\rho$ Mass Measurement Method at the NSCL}
\label{sec:tofbrhonscl}
\begin{figure}[!h]    %{r}{0.6\textwidth}
\begin{center}
  \includegraphics[trim=0 40 0 0,width=0.9\linewidth]{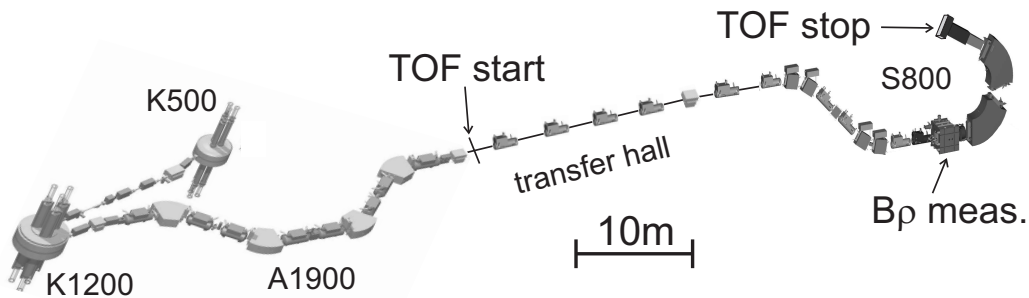}
  \caption{An overview of the facility. The primary relativistic beam is accelerated in the coupled
cyclotrons K500-K1200. Radioactive ions are produced by fragmentation and separated in
the A1900 fragment separator, where the TOF measurements start. The magnetic
rigidity measurements are performed in front of the S800
spectrograph operated with dispersion matched optics. The TOF stop
detector is placed at the S800 focal plane together with an ion
chamber for energy loss measurement and CRDC detectors for particle tracking.}
  \label{NSCLoverview}
\end{center}
\end{figure}

At the NSCL a fast radioactive beam is produced in the A1900
fragment separator~\cite{stolz05}. The 58~m long time-of-flight line
starts at the A1900 extended focal plane and ends at the focal plane
of the S800 spectrograph~\cite{bazin00}, as shown in~Figure~\ref{NSCLoverview}.
\vspace{10mm}
%\vspace{-5mm}
% 
Fast scintillator detectors provide the
timing resolution of $\sigma$=30ps. The relative magnetic rigidity is
measured at~the~momentum dispersive plane by position-sensitive
micro-channel plate detectors~\cite{shapira00} with a position
resolution of $\sigma$=0.3mm. For more details
refer to~\cite{estradenic}.
%The facility was also tested for the isochronous setting option.

\begin{figure}[h] %{r}{0.6\textwidth}
\begin{center}
  \includegraphics[width=0.7\linewidth]{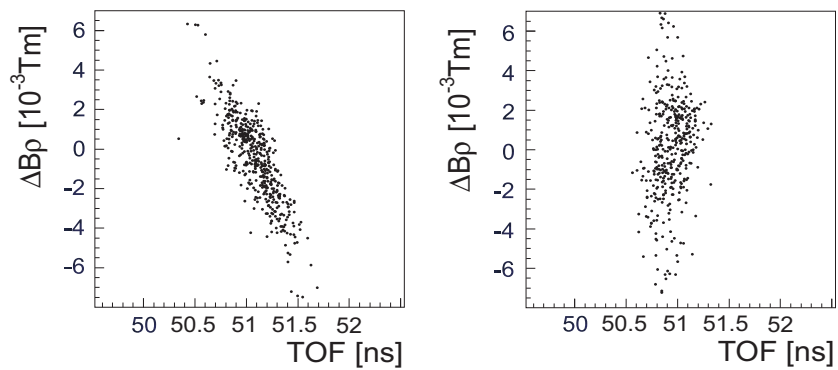}
  \caption{Measured TOF values of $^{79}$Ga particles compared with their magnetic rigidities
  for standard (left panel) and isochronous (right panel) magnetic optics. Projected onto the TOF axis
  the TOF resolution is $\sigma=200$~ps and $\sigma=130$~ps respectively. In the case of standard
   optics the resolution corrected for the magnetic rigidity is $\sigma=80$~ps.}
  \label{iso}
\end{center}
\end{figure}

We also tested the operation of the magnetic optics system in an isochronous mode. 
A~GICO calculation~\cite{GICO} was performed to find an
isochronous magnetic optics setting.
 At the beginning of the experimental
run, the isochronous mode was successfully achieved with a TOF
resolution of $\sigma=130$~ps, an improvement over the
standard optics with $\sigma=200$~ps as shown in Figure~\ref{iso}.
 However, for the TOF corrected with the measured magnetic rigidity  a~resolution of $\sigma=80$~ps is achieved
in the standard dispersion-matched
optics mode.
%Besides this it has also other advantages over the isochronous
%method (see Sec.~\ref{sec:tofmethod}), so 
We therefore decided to use the
TOF-B$\rho$ technique with standard optics.
%
%\begin{wrapfigure}{r}{0.5\textwidth}
%\end{wrapfigure}
%
\section{Mass Measurement of Neutron-Rich $^{86}$Kr Fragmentation Products}
In the first TOF-B$\rho$ mass measurement at the NSCL
a radioactive ion beam in the region of $^{68}$Fe
was produced by fragmentation of a relativistic 100~MeV/u $^{86}$Kr
primary beam on a solid Be target. Production targets with
thicknesses of 46~mg/cm$^2$ and 94~mg/cm$^2$ were alternately used to
populate a wide region of the chart of the nuclides in the
secondary beam without the need for any change in the experimental settings.

\begin{figure}[h] %{r}{0.6\textwidth}
\begin{center}
%a)
  \includegraphics[width=0.9\linewidth]{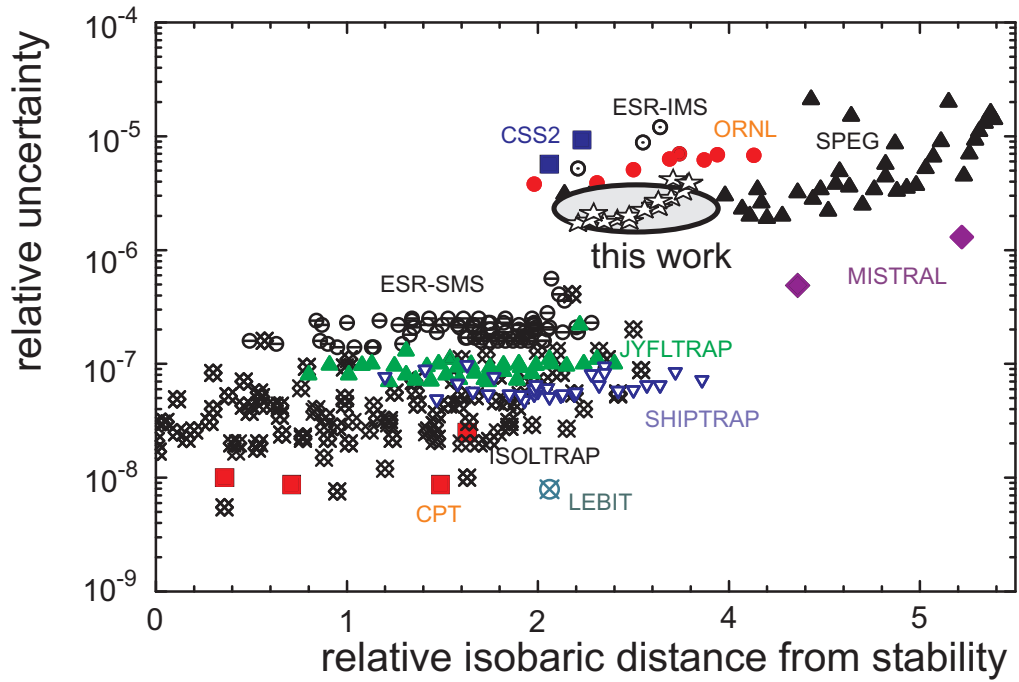}
%\hfill
%  \hspace{5mm}
%b) \hspace{2mm}
%  \includegraphics[width=0.35\linewidth]{figure3.eps}
 \caption{%a) 
Relative experimental uncertainty of mass measurement results versus isobaric
distance from stability for each nuclide obtained from~\cite{lunneyNIC} for comparison with our recent results.
%The programs in brackets correspond to data older than 5 years.
The results from this work are shown with open star symbols.
%}  \caption{
%b) Abundance composition at the moment of charge-particle freeze-out in the high entropy ejecta of
%a~Type II supernova simulation for entropy 100 $k_{B}$ $baryon^{-1}$ and proton-to-nucleon
%ratio $Y_e=0.45$. The reaction network is described in \cite{Far06}.
%The area, covered in this experiment, is marked.
}
%  \label{freezeout}
   \label{lunney}
\end{center}
\end{figure}

The relative mass resolution of $2\times10^{-4}$ obtained in the experiment allowed
 the relative statistical uncertainties to be $(1-3)\times10^{-6}$.
The systematic errors of the order of $(2-3)\times10^{-6}$ in the
preliminary analysis are expected to be reduced in the final results.
Figure \ref{lunney} shows the relative experimental uncertainties of mass measurements
 versus isobaric distance from stability (for details see~\cite{lunneyNIC}).
The results from this work are compared with other mass measurement techniques.
The presented technique has access to more exotic nuclides with the NSCL radioactive beam facility
and is competitive with~SPEG mass measurements.
%\vspace{-3mm}
%
%\vspace{-10mm}

%
\section{Application to Nuclear Astrophysics}
\subsection{Astrophysical R-Process Seed Nuclei}
One of the leading candidates for the location of the r-process 
 is  the high entropy ejecta of Type II supernovae \cite{CTT91,WWH92}.
In this scenario neutrinos not only revive the explosion via heating after the initial shock wave stalls due to the
in-falling material of outer layers, but also drive the material  neutron-rich with
the hardening of the $\bar{\nu_{e}}$ spectra relative to the $\nu_{e}$ spectra at later times.

As this neutron rich material initially consisting of neutrons, protons and $\alpha$-particles
expands adiabatically and cools, heavy particles are formed by a combination of charged-particle
reactions (proton and $\alpha$-captures) and neutron captures. As the temperature decreases,
charged-particle reactions freeze-out and a subsequent neutron capture process proceeds 
if there are enough neutrons remaining.

\begin{figure}[h] %{r}{0.6\textwidth}
\begin{center}
  \includegraphics[width=0.6\linewidth]{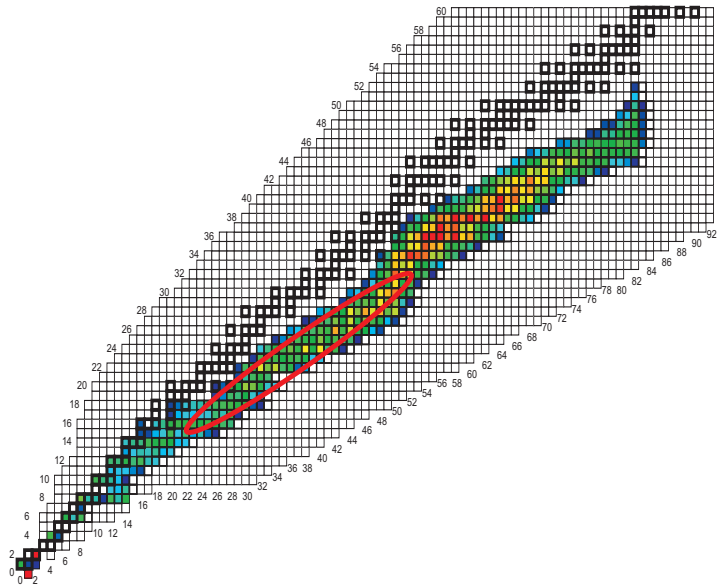}
  \caption{Abundance composition at the moment of charge-particle freeze-out in the high entropy ejecta of
a~Type II supernova simulation for entropy 100 $k_{B}$ $baryon^{-1}$ and proton-to-nucleon
ratio $Y_e=0.45$. The reaction network is described in \cite{Far06}.
The area, covered in this experiment, is marked.}
  \label{freezeout}
\end{center}
\end{figure}

The resulting abundances, when the charge-particle freeze-out occurs,
serve as seeds for the subsequent  neutron capture process \cite{FRT99,Far06}. 
The abundances are
sensitive to neutron separation energies 
not only when neutron captures take over but also to determine 
%(and masses) within this region since they determine 
the initial seed composition. 
Mass values in the required region of nuclides are partially covered  by the experiment discussed in this paper.
Figure~\ref{freezeout} shows the abundance composition at the moment of charge-particle freeze-out
for a case with the entropy of 100 $k_{B}$ $baryon^{-1}$ and the initial proton-to-nucleon ratio
 $Y_e=0.45$.
%As some of these conditions do not evolve into a full r-process, individual neutron capture cross
%sections may also play a role.
%
%
\subsection{Heating of the Accreting Neutron Star Crust and Superburst}
Processes occuring after accretion of matter onto neutron stars in binary systems
have  been related to observables in X-ray binaries.
The ashes of the rp-process sink deeper into the crust and
the electron chemical potential $\mu_{e}$ rises with the increasing density.

A calculation, seeding on a single beta-stable even-even rp-process nuclide,
was performed in \cite{haenselzdunik1,haenselzdunik2}.
Two processes were the primary deep crustal heating sources; pycnonuclear fusion at $\rho > 10^{12}$g/cm$^2$,
and 
two-stage electron-capture at densities $\rho > 10^{11}$g/cm$^2$
that includes  cold n-emission to maintain mechanical equilibrium within a Wigner-Seitz cell
beyond cold neutron drip density.
%and  pycnonuclear fusion at $\rho > 10^{12}$g/cm$^2$ were the primary deep crustal heating sources.

New X-ray observations have led to the discovery of superbursts that are
roughly 10$^3$ times more energetic compared to Type I X-ray bursts \cite{Kuulkers2003}.
The thermally unstable ignition of $^{12}$C at $\rho > 10^{9}$g/cm$^2$
 was a proposed  scenario for superbursts \cite{superburst1,superburst2}, although
the heat production in the deep crust
 would not be sufficient to explain observed superburst recurrence timescales of less than
10 years~\cite{Wijnands2001, Kuulkers2004}.

The recent calculations of crustal heating using a realistic mix of rp-ashes~\cite{gupta}
and including electron capture into excited states followed by a radiative deexcitation
predict increased heating and therefore shallower superburst ignition and
lower recurrence times between 2--3 years.
%of just below 2 years.
%The nuclear model that was used to calculate the  Gamow-Teller
%transition strength functions is described in
%detail in \cite{Moller1990Newdevelopment}. Neutrino loss rates were accurately calculated
%for each transition, instead of assuming (as in \cite{haenselzdunik1,haenselzdunik2})
%that a fixed fraction of $(\mu_{e}- E_{thresh})$ is lost to neutrinos.

The results are very sensitive to the nuclear physics of
neutron-rich nuclei up to A=106. In particular masses are needed for
electron capture thresholds, heating rates from excited state captures, and the separation
energies playing a crucial role in neutron reaction rates. 
%This sensitivity is a consequence
%of the pronounced shell and sub-shell structure of neutron-rich nuclei, which can lead to drastic
%changes in shape, single-particle level structure, and electron capture strength distributions,
%even between nuclei of similar $Z, A$. 
New mass measurements  in this region, including the experiment discussed in this paper, are important for the calculation
to replace the  FRDM mass model values \cite{moller} that were used.

%Missing experimental mass values should be covered with experimental data including the results
%from the experiment discussed in this paper.
%In the recent calculation the  FRDM mass model values \cite{moller} were used.
%New mass measurements are important for the calculation to replace
%the  FRDM mass model values \cite{moller} that were used because
%of missing experimental data.

\section{Conclusions}
 The mass values in the region covered by this work
are crucial for astrophysical calculations such as
the investigation of the thermal balance in the crust of
accreting neutron stars or the calculation of r-process seeding nuclei.

Experimental masses for neutron-rich nuclides have often been shown to be
unreliable \cite{matosexon1, matosexon2} and new mass measurements are required.

The technique presented in this work is expected to enable 
the measurements of nuclides far from stability with relative
uncertainties of the order of $2\times10^{-6}$. This method is now established at the NSCL
coupled-cyclotron facility and will be used to measure masses
of numerous very exotic nuclides.

\end{document}